\documentclass{aa}
\usepackage{graphicx}

\def\gtrsim{\mathrel{\hbox{\rlap{\hbox{\lower4pt\hbox{$\sim$}}}\hbox{$>$}}}}
\def\ltsim{\mathrel{\hbox{\rlap{\hbox{\lower4pt\hbox{$\sim$}}}\hbox{$<$}}}}

\begin{document}

\title{EK Eridani: the tip of the iceberg of giants which have evolved from  magnetic Ap 
stars\thanks{Based on data obtained using the T\'elescope Bernard Lyot at Observatoire du 
Pic du Midi, CNRS/INSU and Universit\'e de Toulouse, France.}}

\author{M. Auri\`ere\inst{1}, R. Konstantinova-Antova\inst{2},  P. Petit\inst{1}, 
C. Charbonnel\inst{3,1}, B. Dintrans\inst{1}, F. Ligni\`eres\inst{1}, T. Roudier\inst{1}, 
E. Alecian\inst{4},  J.F. Donati\inst{1},  J.D. Landstreet\inst{5},   G.A. Wade\inst{4}}
\offprints{michel.auriere@ast.obs-mip.fr}
\institute{Laboratoire d'Astrophysique de Toulouse- Tarbes, Universit\'e de Toulouse, CNRS, 
57 Avenue d'Azereix, 65008 Tarbes, France
 \and
Institute of Astronomy, Bulgarian Academy of Sciences, 72 Tsarigradsko shose, 1784 Sofia, Bulgaria
 \and
Geneva Observatory, University of Geneva, 51 Chemin des Maillettes, 1290 Versoix, Switzerland
 \and
Department of Physics, Royal Military College of Canada,
  PO Box 17000, Station 'Forces', Kingston, Ontario, Canada K7K 4B4
 \and
Department of Physics \& Astronomy, The University of Western Ontario, London, Ontario, Canada, N6A 3K7}

 \date{Received ??; accepted ??}

\abstract
 {}
{We observe the slowly-rotating, active, single giant, EK Eri, to study and infer the nature of its magnetic field directly.}
{We used the spectropolarimeter NARVAL at the Telescope Bernard Lyot, Pic du Midi Observatory, 
and the Least Square Deconvolution  method to create high signal-to-noise ratio Stokes V profiles. 
We fitted the Stokes V profiles with a model of the large-scale magnetic field. We 
studied the classical activity indicators, the Ca~{\sc ii} H and K lines, the Ca~{\sc ii} infrared 
triplet, and $H\alpha$ line.}
{We  detected the Stokes V signal of EK Eri securely and measured the longitudinal magnetic 
field B$_l$ for seven individual dates spanning 60\% of the rotational period. The measured longitudinal magnetic 
field of EK Eri reached about 100 G and was as strong as fields observed in RSCVn or FK Com type stars: this was found to be  
extraordinary when compared with the weak fields observed at the surfaces of slowly-rotating MS 
stars or any single red giant previously observed with NARVAL. From our modeling, we infer that 
the mean surface magnetic field is about 270 G, and that the large scale magnetic field is dominated 
by a poloidal component. This is compatible with expectations for the descendant of a 
strongly magnetic Ap star.}
{}

   \keywords{stars: individual: EK Eri -- stars: magnetic fields -- stars: late type}
   \authorrunning {M. Auri\`ere et al.}
   \titlerunning {Direct detection of a magnetic field in EK Eri}

\maketitle

\section{Introduction}

EK Eri (HR 1362, HD 27536) is the most slowly rotating active giant known (Strassmeier et al. 1990a, 1999). These authors remarked that, because of its long 
rotation period (306.9 d), its magnetic activity could not be maintained by a classical, solar-type, dynamo. Stepie\'n (1993), supported by 
Strassmeier et al. (1999), proposed that EK Eri was the descendant of a strongly magnetic Ap star. A large scale fossil magnetic field interacting 
with a newly-born, deepening, convection zone could therefore be the origin of its activity.  

It is generally supposed that deep convective zones can shred large-scale magnetic 
fields; the most well studied phase, in this respect, is the pre-main sequence phase (Moss, 2003). Large scale fossil magnetic fields are believed to be present in some pre-main sequence stars (Alecian et al. 2008), some main-sequence stars (Landstreet, 1992), and in some white dwarfs and neutron stars (Wickramasinghe et al. 
2005, Ferrario et al., 2006), which suggests that these fields are able to survive the red giant phase. 

 Strong large-scale magnetic fields (hosted by descendants of Ap stars) were 
invoked by Charbonnel \& Zahn (2007b) to inhibit thermohaline mixing and explain the deviant composition of some giants (Charbonnel \& Do Nascimento 1998, 
Charbonnel \& Zahn 2007a).

In this context, EK Eri appears to be a promising object for understanding the interaction of a large-scale magnetic field and deep convection zone. We therefore used spectropolarimetry to study directly the surface magnetic field of EK Eri. In Sect. 2 of this paper, we describe our observations with NARVAL and  our direct results. In Sect. 3, we discuss the fundamental parameters of EK Eri and its outstanding activity; we used standard models to determine the evolutionary status of the star. In Sect. 4 we study the star's observed magnetic field and its possible origin. We summarise our conclusions in Sect. 5.

\section{Observations}

\subsection{Observations with NARVAL}

The observations of EK Eri were obtained at the 2m Telescope Bernard Lyot (TBL) of Pic du 
Midi using NARVAL, a new generation spectropolarimeter (Auri\`ere 2003). NARVAL is a copy of 
ESPaDOnS, installed at CFHT, towards the end of 2004
(Donati et al. 2006a, 2008). NARVAL consists of a Cassegrain polarimetric module and a fiber--fed echelle spectrometer
allowing the entire (polarimetrically analysed) spectrum from 370 nm to 1000 nm to be recorded in each
exposure. The resulting  40 orders are aligned on the CCD frame by 2 cross-disperser prisms.
NARVAL was used in polarimetric mode with a spectral resolution of 
about 65000. Stokes I (unpolarized) and Stokes V (circular polarization)
parameters were obtained by means of 4 sub-exposures between which the 
retarders (Fresnel rhombs) were rotated to exchange the beams 
in the entire instrument and reduce spurious polarization signatures. 

During the technical tests and science demonstration time, magnetic and non-magnetic stars were observed.
These data indicated that NARVAL is 30 times more sensitive than its predecessor, MuSiCoS 
(Baudrand \& Boehm 1992, Donati et al. 1999).  We observed EK Eri on 7 different dates between  September 2007 and April 2008.  For these dates, Table 1 provides the rotational phase (based on a rotation period of 306.9 d, the origin of which is our September 2007 observation i.e. HJD= 2454364.6925), the total exposure time, and  the maximum S/N ratio per spectral bin (of 2.6 km s$^{-1}$ size) in Stokes I. 
The extraction of the spectra was achieved using Libre-ESpRIT (Donati et al. 1997a, 2008),
a fully automatic reduction package installed at the TBL. 

To complete the Zeeman analysis, Least-Square Deconvolution 
(LSD, Donati et al. 1997a) was applied to all observations. We used a mask calculated for an effective 
temperature of $5250 °K$, $\log g =3.0$, and a microturbulence of 2.0 km s$^{-1}$, consistent with physical parameters given by Dall et al. (2005). In the present case this method enabled us to average about 11,000 lines and to derive
Stokes V profiles with  of S/N ratio that improved by a factor of about 40 in comparison with that for single lines.  We 
then computed the longitudinal magnetic field B$_{l}$ in G, using the 
first-order moment method (Rees \& Semel 1979, Donati et al. 1997a). 

The activity of the star during the same nights was monitored by using line--activity indicators. We employed 
measurements of the intensity ratio I/I(395 nm) for the Ca~{\sc ii} K (its V and R components,where I(395) is the continuum measurement), 
and the relative intensity with respect to the continuum (R$_c$) of the Ca~{\sc ii} IR triplet (854.2 nm component) and  
H$_\alpha$ (Strassmeier et al. 1990b). 

\begin{table*}
\caption{Observations of EK Eri (for details, see sect. 2.1): Dates, rotational phases, exposure times, maximum S/N in Stokes I, I/I(395) values for Ca~{\sc ii} K, R$_c$ values for Ca~{\sc ii} 854.2 and $H\alpha$, B$_l$ values and their uncertainties in G for the magnetic field and radial velocity.}           
\centering                         
\begin{tabular}{c c c c c c c c c c c}     
\hline\hline               
Date      &Rot. & Exp.& S/N&Ca~{\sc ii} K & Ca~{\sc ii} K &Ca~{\sc ii} &$H\alpha$& B$_l$& $\sigma$& $V_{rad}$  \\
          &Phase& sec.&    &V comp.& R comp.&854.2&         & G    &  G      & km s$^{-1}$   \\
\hline                        
20 Sep 07 & 0.00& 3600& 816& 0.55  & 0.53   &0.319& 0.226   &-98.6 & 1.0     & 7.118  \\
12 Nov 07 & 0.17& 2280& 900& 0.49  & 0.46   &0.291& 0.233   &-21.1 & 0.7     & 7.047  \\
13 Nov 07 & 0.17& 2280& 731& 0.50  & 0.47   &0.297& 0.233   &-21.4 & 0.9     & 7.053  \\
19 Jan 08 & 0.39& 2400& 545& 0.47  & 0.43   &0.303& 0.230   &-28.9 & 1.2     & 7.096  \\
20 Jan 08 & 0.39& 2400& 720& 0.47  & 0.43   &0.295& 0.227   &-31.6 & 0.9     & 7.110  \\
06 Feb 08 & 0.45& 2400& 779& 0.47  & 0.44   &0.302& 0.234   &-45.6 & 0.8     & 7.111  \\
03 Apr 08 & 0.63& 1200& 675& 0.50  & 0.48   &0.339& 0.250   &-62.7 & 1.2     & 7.145  \\
\hline                                 
\end{tabular}
\end{table*}

\begin{figure}
\centering
\includegraphics[width=8 cm,angle=0] {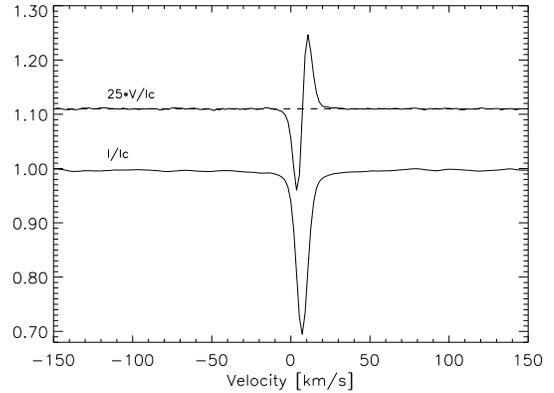} 

\caption{LSD profiles of EK Eri as observed on 20 Sep. 2007 
with NARVAL. From bottom to top, Stokes $I$ and 
Stokes $V$ are presented. For display purposes, the profiles are shifted vertically, and  the Stokes $V$ profile is expanded by a factor of 25. The dashed 
line illustrates the zero level for the Stokes $V$ profile.}
\end{figure}

\begin{figure}
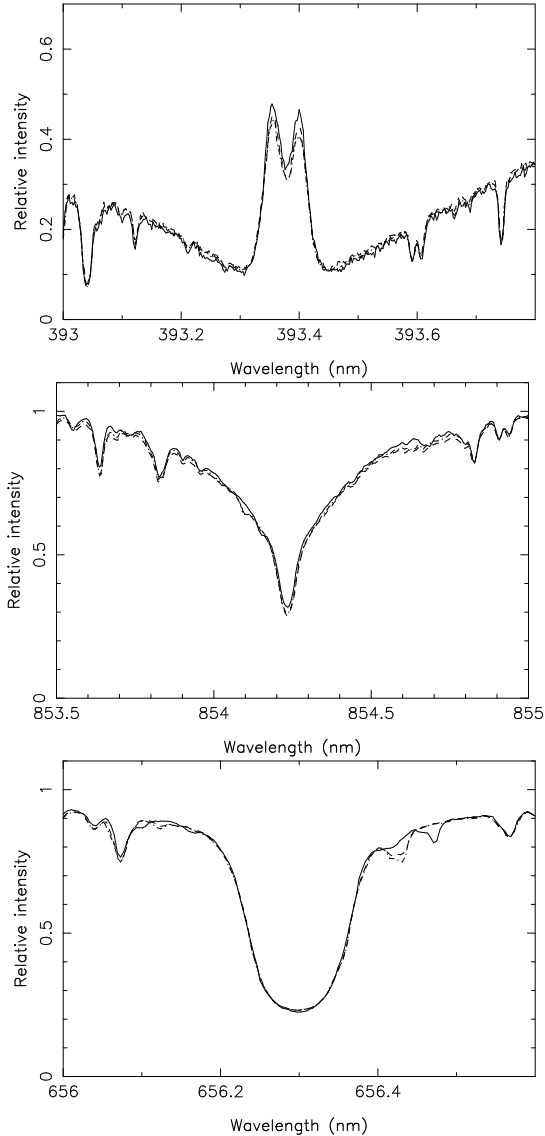

\centering
\includegraphics[width=5 cm,angle=270] {0502fig2a.ps} 
\includegraphics[width=5 cm,angle=270] {0502fig2b.ps}
\includegraphics[width=5 cm,angle=270] {0502fig2c.ps}
           
\caption{Profiles of Ca~{\sc ii} K, Ca~{\sc ii} IR 8542, H$\alpha$ for Sep. 2007 and Nov. 2007 observations of EK Eri; between these two dates,  the greatest variation 
in B$_l$ is observed. The changes in the  $H\alpha$ wings are due to telluric lines.}
\end{figure}

\subsection{Results of observations}

A significant Zeeman Stokes V signal (ranging between $10^{-3}$ - 6 x $10^{-3}$ of the continuum) was detected in each observation. Figure 1 shows the LSD spectra obtained for EK Eri on  20 Sep. 2007.
Table 1 reports the 7 measurements of longitudinal magnetic field B$_{l}$, as well as the line activity indicators.  The two magnetic observations on 12 and 13 Nov. 2007, 
and the two on 19 and 20 Jan. 2008, are similar, which is consistent with a slowly rotating star.  

The H \& K Ca~{\sc ii} emission profiles are double-peaked and 
asymmetric, the short wavelength peak being stronger than the longward peak as shown (Fig. 2 for the K line). This asymmetry is in the same sense as for the 
integrated disk of the Sun, and for most red giants detected by the ROSAT satellite (Smith \& Shetrone 2000), and indicates vertical motions in the chromosphere. Table 1 and Fig. 3 
show that at phases 0  and 0.6, $|B_l|$ (the unsigned, longitudinal, magnetic field) and the Ca~{\sc ii} activity indicators reach local maxima. However, the correlation with $|B_l|$ is  not extremely tight; the Ca~{\sc ii} activity indicators vary more smoothly. A close correlation is not expected between  $|B_l|$ and activity indicators, if the photospheric magnetic structure is complex and nearby opposite-polarity magnetic regions can cancel altogether. If 
confirmed, a correlation between $|B_l|$ and the Ca~{\sc ii} activity indicators could suggest that the magnetic structure is a large-scale one, instead of consisting of small magnetic 
areas of opposite polarities.

$H\alpha$ profiles appear as normal absorption features (Eaton, 1995) and exhibit only marginal variability for all of our observations
(Fig. 3).  This indicates that we do not observe an emission component from the chromosphere in this line. 
A similar $H\alpha$ feature was reported for other active single giants by Fekel \& Balachandran (1993).

The last column of Table 1 provides our radial velocity measurements of EK Eri measured on the LSD Stokes I profiles from Gaussian fits. The accuracy of ESPaDoNS and NARVAL radial velocities is about 20-30 m s$^{-1}$ (Moutou et al. 2007). The observed variations of about 100 m s$^{-1}$ in the amplitude of $ V_{rad} $ are therefore significant. Figure 4 shows that the variations of $ V_{rad} $ are well correlated with  $|B_l|$, which supports the proposal of Dall et al. (2005) that these variations are due to activity variations.

\begin{figure}
\centering
\includegraphics[width=8 cm,angle=0] {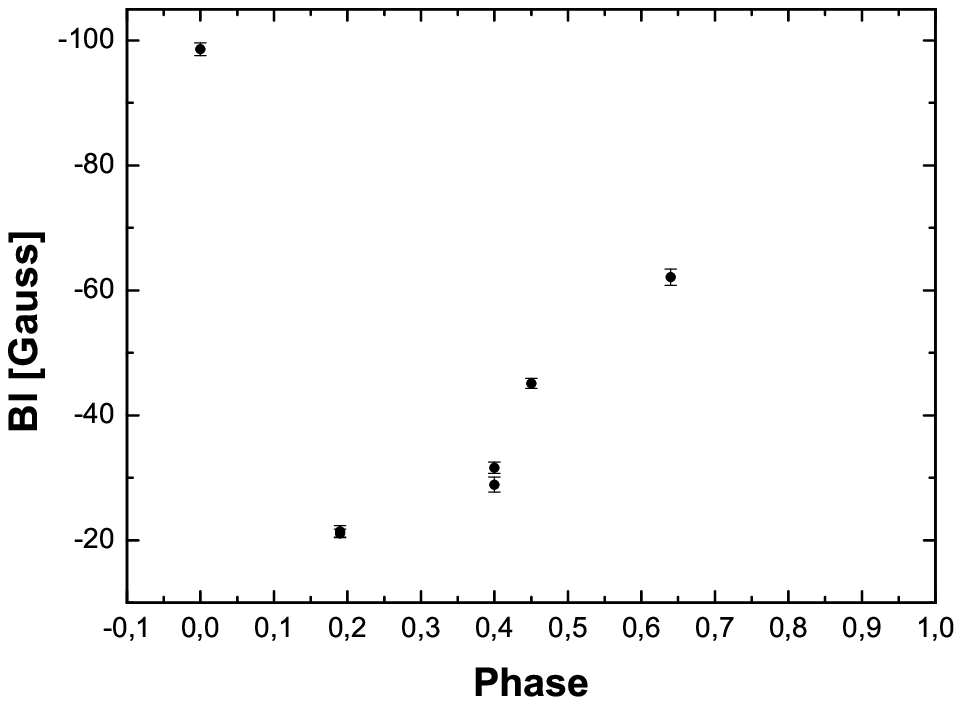} 
\includegraphics[width=8 cm,angle=0] {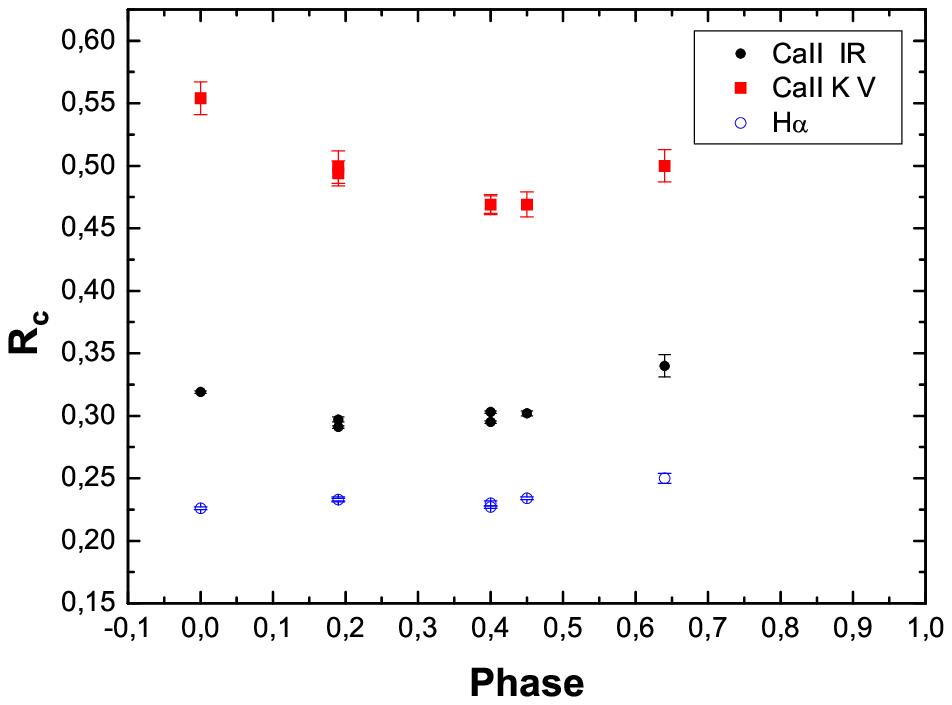}

\caption{Longitudinal magnetic field (upper panel) and behavior of activity indicators (lower panel: from top to bottom: Ca~{\sc ii} K, Ca~{\sc ii} IR 854.2, H$\alpha$) for EK Eri with respect to P=306.9 d phase (see Table 1).}
\end{figure}

\begin{figure}
\centering
\includegraphics[width=8 cm,angle=0] {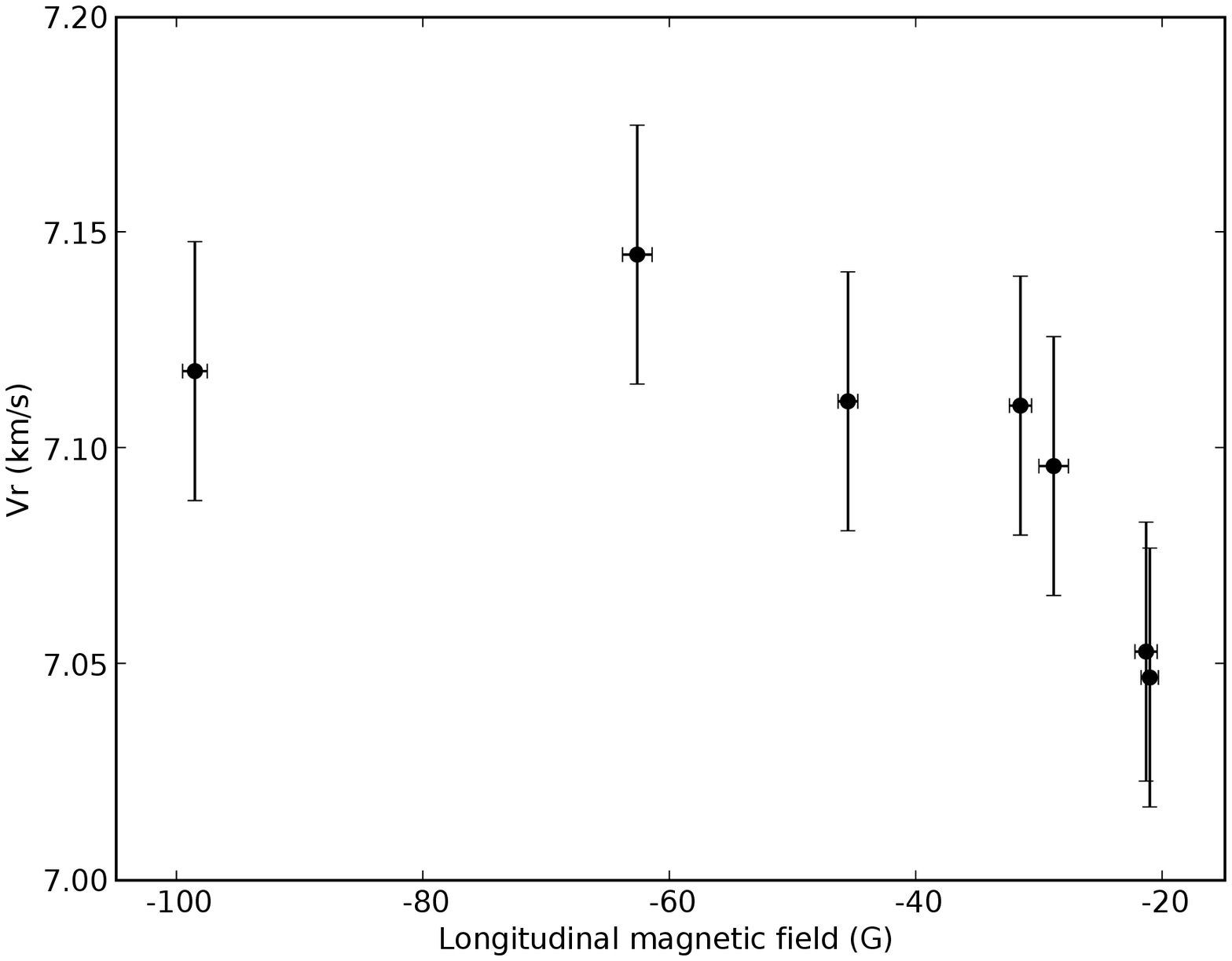} 

\caption {Radial velocity (ordinates) versus longitudinal magnetic field (X-axis) for EK Eri. Error bars are indicated.}
\end{figure}

\section{Rotation, evolutionary status and activity of EK Eri}

EK Eri is considered to be exceptional since it is an active red giant with a long rotational period, and it violates strongly the classical relations 
between rotation period and activity indicators values observed for stars of its class. We therefore examine its rotational and evolutionary status 
before dealing with the interpretation of our magnetic field measurements.

\subsection {The rotational period of EK Eri}

EK Eri has long been known to be a variable star (Rufener and Bartholdi 1982). Strassmeier et al. (1999) presented results of a photometric investigation 
using data obtained over twenty years, involving about ten telescopes and two thousand measurements. These data were processed by a 
least squares periodogram analysis, which measured a single, strong, period of $306.9\pm0.4$ days (Strassmeier et al. 1999). Studying two subsamples, each 2500 days long, they measured periods instead of $315.5 \pm 5$ days and $294 \pm 2$ days.  No indication 
for a significantly shorter period is found (Strassmeier et al. 1990, 1999). Hipparcos measurements did not reveal  the periodic nature of the stellar variability, but did provide evidence of a maximum frequency peak at about 340 days (Koen \& Eyer 2002).  

This period could be due to spots on two or three active longitudes, as thought to appear 
on Arcturus (Gray \& Brown 2006); in this case, the period found from modulation of these features will be one-half or one-third of the rotation period, which  would therefore
be 2 or 3 times the (306.9 d) photometric value, and EK Eri would appear even more of an exceptional case. The period could also be due  to an asymmetric polar spot observed on a rapid rotator observed pole-on, and one might have slow polar rotation while equatorial rotation could be more rapid. However, because of the large amplitude in both photometric and $B_l$ variations, a rather large inclination is expected.

Our observations indicate that the magnetic field varies significantly with  phase.   We also confirm variations in $V_{rad}$ of amplitude reaching about 100 m s$^{-1}$ as observed by Dall et al. (2005) with HARPS. Our  modeling of the large-scale magnetic field,  presented in Sect. 4, and radial velocity measurements, will therefore constrain the geometry of the star and its rotational period as soon as it has been observed during more than a complete rotation.

\subsection{Evolutionary status of EK Eri}

It is important to assess the evolutionary status of EK Eri because HD181943, another long period (385 days, Hooten \& Hall 1990), active, giant,
 Ap star descendant candidate (Stepie\'n, 1993), was found to have a photometric period that is far longer (about 2000 days, Fekel et al. 1995), and not equal to its rotational 
period, and could  be a K1V-IV star or even a pre-main-sequence star (Strassmeier 1991, Strassmeier et al. 1999).

Figure~5 shows the position of EK Eri in the HR diagram (data from Strassmeier et al. 1999).
Standard, non-rotating, evolutionary tracks for different stellar masses computed for solar metallicity 
with Asplund et al.~(2005) chemical composition and  mixing length parameter 
equal to 1.6 are also shown (Charbonnel et al. in preparation). 
The position of EK Eri in this diagram indicates an initial mass of 2$\pm 0.1$ M$_{\odot}$. 
The star is located at the end of the Hertzsprung gap,
as previously asserted by Strassmeier et al.~(1999). 
At the precise location of EK Eri, our 2~M$_{\odot}$ model has a radius of 4.68~R$_{\odot}$; 
this value is in excellent agreement with that derived from the Hipparcos parallax by means of the 
bolometric-luminosity -- effective-temperature relation (i.e., 4.7$\pm$0.3~R$_{\odot}$, 
Strassmeier et al.~1999). 

At this evolutionary point, the star experiences the so-called first dredge-up, and 
the convective envelope of our 2~M$_{\odot}$ model contains $\sim 0.37$ M$_{\odot}$, 
(which will engulf $\sim 1.7$ M$_{\odot}$ at its maximum extent, at the end of the first 
dredge-up, and later on the red giant branch).
As a consequence, the surface lithium abundance starts to decrease due to dilution 
by Li-free regions. 

 We note that the observed surface Li abundance of EK Eri (Strassmeier et al. 1999, Dall et al. 2005) is 
between the theoretical predictions for the 1.8 and 2.2~M$_{\odot}$ rotating models
of Palacios et al.~(2003).

It is now possible to infer the main-sequence progenitor of EK Eri. A 2~M$_{\odot}$ MS star is an A5V star (Allen, 2000). These stars are in general quite rapid rotators,  with equatorial velocities higher than 100 km s$^{-1}$. The standard model, which includes the transport of angular momentum (Zahn, 1992), predicts that a 2~M$_{\odot}$ main sequence star with $v_{rot} = 110$ km s$^{-1}$ would  rotate at $v_{rot} = 17$ km s$^{-1}$ in the evolutionary state of EK Eri, while $v_{rot} = 0.13$ km s$^{-1}$ is inferred using a 306.9 d rotational period. This strongly suggests that the progenitor of EK Eri was a slowly rotating A-type star. Such slow rotators are mainly chemically peculiar (CP)  stars (Abt and Morrell, 1995). About  half of CP A-type stars are magnetic Ap stars (Abt and Morrell, 1995), which supports the proposal by Stepie\'n (1993) of the  progenitor of EK Eri. Dall et al. (2005) measured a normal abundancy of Cr in EK Eri. This element would have been significantly over-abundant if it had been a main-sequence A5 Ap star, but  would have been diluted during the G phase of the dredge-up (Bohm-Vitense, 1993).

\begin{figure}
\centering
\includegraphics[width=8 cm,angle=0] {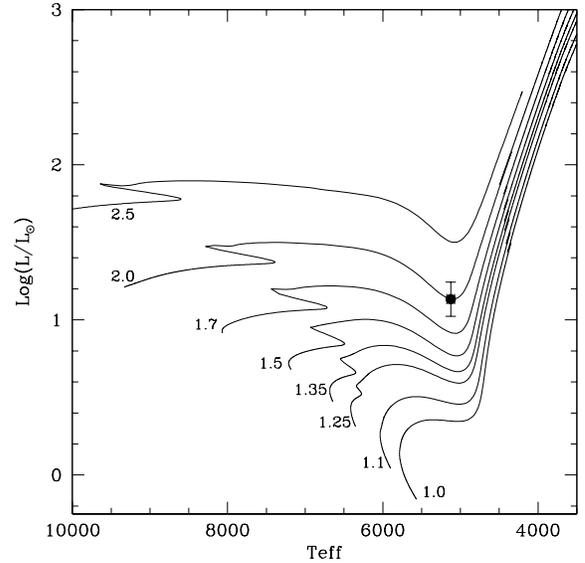} 

\caption{HR diagram showing the position of EK Eri (observational values 
by Strassmeier et al.~1999; the size of the symbol indicates the uncertainty of Teff) 
with respect to standard evolutionary tracks for solar composition. 
Masses are from 1.0~M$\sun$ up to 2.5~M$\sun$ as indicated. 
EK Eri lies on the 2~M$_{\odot}$ evolutionary track. 
For details see sect. 3.2.}
\end{figure}

\subsection {The activity of EK Eri}

The two main proxies used to monitor stellar magnetic activity are the intensity of the Ca~{\sc ii} H \& K lines (chromospheric activity), and X-ray emission 
(coronal activity). High magnetic activity in late-type stars may also cause photometric variability due to spots (Berdyugina, 2005). The relation between 
Ca~{\sc ii} K\& H line intensity and the rotation rate of evolved stars has been well documented: the correlation is so significant that Young et al. (1989) predicted 
the rotation period of active, single giants on the basis of their Ca~{\sc ii} H \& K fluxes. Strassmeier et al. (1994) confirmed that the Ca~{\sc ii} surface fluxes from evolved 
stars scaled linearly with stellar rotational velocity. Although a significantly large scatter exists, EK Eri is the only known outlier of this 
relation (with HD181943, but see Sect. 3.2.): its activity is similar to that of the single giants with rotation periods close to 10 days studied by  Strassmeier et al. (1994). Gondoin (2005) studied the relation between X-ray activity and 
rotation in G giants: their X-ray surface flux was found to increase linearly with the angular rotation velocity. EK Eri has an X-ray luminosity of about $10^{30}$ erg/s (H\"unsch et al. 1998). 
 It is as  bright as the 10 outstanding X-ray objects in a list of about 90 G-K giants observed by Einstein or ROSAT and studied by Gondoin (1999).

EK Eri is also an interesting photometric variable star: its photometric period appears to have been constant for about 20 years, 
but the amplitude of the variations is variable and has reached about 0.2 mag in V and 0.05 in B-V (Strassmeier et al. 1999).

\section {The magnetic field of EK Eri}

\subsection {Observed properties of the magnetic field of EK Eri}

Figure 1 and 6 show that the Stokes V profile of EK Eri harbours a significant Zeeman signature in all of our observations. The simple shape  is unsurprising since the small $v \sin i$ value of about 1 km s$^{-1}$ (Dall et al. 2005) does not enable detection of possible complicated 
spot structures  with NARVAL's spectral resolution. The mean longitudinal magnetic-field measurements presented in Table 1 indicate that B$_l$ can reach 
about 100 G, which is the level observed in the case of rapidly rotating giants, such as the binary RS CVn  star II Peg (MuSiCoS observations, Petit et al., in 
preparation) or the single FK Com giant HD 199178 (Petit et al. 2004). The main-sequence slow rotators with active chromospheres that have been  studied up to now, for example $\xi$ Boo A (Petit et al. 2005) or solar twins (Petit et al. 2008), have weak B$_l$ values, between a few G and about 20 G. A dozen other single giant stars were also observed with NARVAL, for which magnetic fields were detected directly and B$_l$ measured. The maximum unsigned $B_l + 3\sigma$ observed for the entire sample was 15 G for V390 Aur (Konstantinova- Antova et al. 2008), 
whereas this quantity reaches about 100 G 
for EK Eri. V390 Aur has a rotation period of 9.8 d, and EK Eri reached the same activity level (in Ca~{\sc ii} K and X-rays emission). Our direct detection and measurements of the surface magnetic field therefore strengthened the result obtained with chromospheric and coronal proxies, i.e. 
the strength of the large-scale magnetic field of EK Eri was similar to that detected for very rapidly rotating giant stars.

\subsection {Modeling the Stokes V time series}

The 7 spectra at our disposal  (observed at 5 different rotation phases)
correspond to slightly more than half a rotation
period of EK Eri (assuming a 306.9 d rotation period). With the hypothesis
that all variation that we observe in the Stokes V LSD profiles
originates in rotational modulation, we use this initial data set to
reconstruct the surface magnetic geometry of the star with the
Zeeman-Doppler Imaging inversion method (Donati et al. 1997b), bearing in
mind that the sparse phase sampling cannot offer more than a rough view of
the magnetic geometry of the star. Using spherical harmonics
to model the magnetic topology, we split the magnetic field into a
poloidal and toroidal component (Donati et al. 2006b). Apart from the
rotation period, the stellar model that we use for profile synthesis includes a
$v \sin i$ of 1~km s$^{-1}$, an inclination angle of 80$^\circ$, and a linear limb
darkening coefficient equal to 0.75. We limit the spherical harmonics
expansion to $\ell \le 3$.

The modeled time-series of Stokes V profiles is compared with
observations in Fig. 6. The $\chi2$ that we reach is not of higher quality than 7, mostly
because our line model does not include any Stokes V asymmetry, while this
asymmetry shows up in EK Eri Stokes V profiles, similarly to other cool
stars (Petit et al. 2005) and to solar magnetic elements (see e.g. Solanki
1993). In the case of EK Eri  we note that the blue lobes of the profiles
are always deeper than the red ones, with a difference in amplitude of
about $3\times10^{-4}I_c$, where $I_c$ is the continuum level. We note
that the  $\chi2$ is not improved by increasing the $\ell$ limit of our
model, and is unchanged when the maximum $\ell$ is reduced to
2.

The inferred mean surface magnetic field is about 270 G, with a
large-scale magnetic field dominated by the  poloidal component (the
toroidal component contains less than one percent of the reconstructed
magnetic energy and the magnetic topology remains unchanged if we assume a
purely poloidal field in the inversion process). The largest fraction of
the magnetic energy is contained in the dipole and quadrupole terms, which
represent 96 per cent of the energy. The magnetic-field
distribution is also mostly axisymmetric, and 85 per cent of the magnetic
energy are present in modes with $m = 0$.

These magnetic quantities are not significantly affected when we vary
$v \sin i$ between 0.5 and  1.5~km s$^{-1}$ (the range suggested by Dall et al. 2005) and the inclination angle between 60$^\circ$
and 90$^\circ$. Because of the spectral resolution of HARPS, 
about 100000 (or NARVAL, about 65000) $v \sin i = 0$ km s$^{-1}$ cannot be totally excluded. However, this would correspond to a pole-on observation, and as we argue in Sect. 3.1, we expect a rather large inclination for EK Eri, which explains our choice of parameters. The main difference between the various possible maps is
observed in the energy division between the dipolar and quadrupolar
magnetic components. While an inclination angle close to 80$^\circ$
produces a mainly quadrupolar topology (60 per cent of the energy in the 
$\ell = 2$ modes), the magnetic field becomes a mainly dipolar
distribution (93 per cent of the energy in the  $\ell = 1$ modes) in the case of
$i=60^\circ$. Fitting Stokes V profiles using data for a full rotation of the star should enable us to more fully constrain the inclination and topology of EK Eri.

\begin{figure}
\centering
\includegraphics[width=6 cm,angle=0] {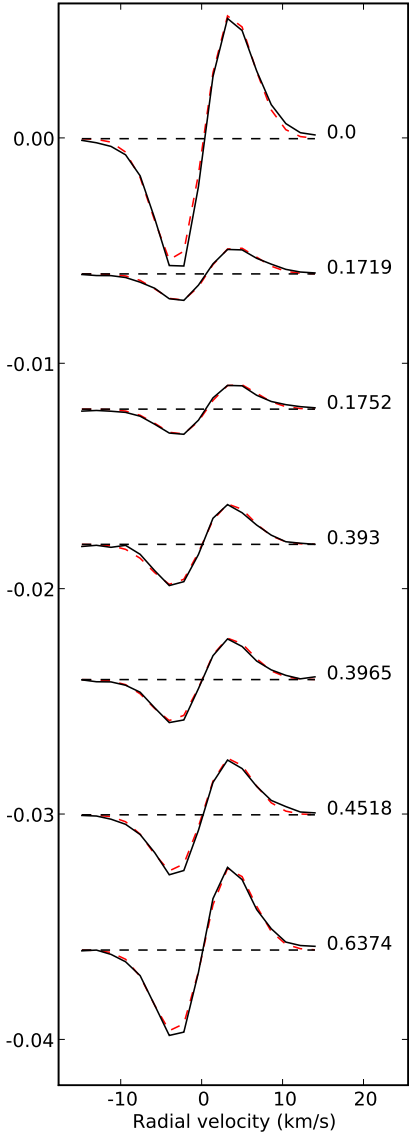} 

\caption{Fit of our Stokes V LSD profiles (black full line) with one of the models described in Sect. 4.2 (red dashed line). Rotational phases are indicated on the right of the profiles. The dashed lines illustrate the zero level. For display purpose, the profiles are shifted vertically.}
\end{figure}

\subsection {Possible origin of the magnetic field in EK Eri}

In the classical (solar-type) dynamo context (Parker 1977), Durney \& Latour (1978) 
showed that the level of activity should be a function of the rotation 
period, $P_{rot}$, divided by a turnover convective timescale $\tau_c$. 
Qualitatively, the Rossby number ($ R_0 = P_{rot} / \tau_c$) must be smaller than 1 
for the successful operation of a solar-type dynamo. 
 
For giants at the first dredge-up, $\tau_c$ (when it is the longest) was estimated 
by Gunn et al. (1998) to be less than about 130 days for stars of  masses smaller than 
2.2~M$_{\odot}$. 
From his own computations, Gondoin (2005) found that the longest $\tau_c$ measured for 
his giant sample was 120 days. 
In our 2.0~M$_{\odot}$ model at the location of EK Eri, the value of $\tau_c$ at the 
base of the convective envelope was approximately of 150 d. 
As a consequence, 
the 306.9 d period of EK Eri is inconsistent with a $ R_0 $ smaller than 1. 
The activity of EK Eri is therefore incompatible with the classical 
solar-dynamo requirements. 
It is generally believed that evolved red giants of long rotation period 
should host only  small surface magnetic fields, which appeared to be 
confirmed by observations (Landstreet, 2004). However, the formation of planetary nebulae 
may require that magnetic fields are produced in  AGB stars, and some 
models have been proposed for this purpose. Blackman et al. (2001) and Dorch (2004) 
proposed models that can provide surface fields of several hundred 
of G at the surface of slow rotators. But if such strong surface magnetic fields were generally present in slowly 
rotating giants, the level of activity of EK Eri would not be so exceptional.

In this context the proposal of Stepie\'n (1993) and Strassmeier et al. (1999) that the magnetic field in EK Eri is of fossil origin, i.e. originating in  an 
Ap star parent, is worth exploring further.  The mean surface magnetic field inferred by our modeling presented in Sect. 4.2 is 270 G: this is of the same order of magnitude as the 
predicted surface magnetic field of the descendant of an Ap star similar to the strong-field star 53 Cam, proposed by Stepie\'n (1993). The estimate of Stepie\'n (1993) relied on the dilution of Ap-type magnetic field during the expansion to the subgiant phase. Now we must study the interplay between the convective envelope and a pre-existing large-scale magnetic field. This issue is important for accounting for the active phenomena in the chromosphere (line-activity indicators) and in the corona (X-ray emission). For example, Soker and Zoabi (2002) showed that turbulent 
dynamo of $\alpha^2$-type  may amplify a large-scale magnetic field (e.g. Brandenburg, 2001). However, since the seed field is weak in this model, the strength of the resultant magnetic 
field is moderate. If the initial large-scale magnetic field is as strong as the remnant of an Ap star one, the observations of 
EK Eri could be explained naturally.

Now, the result of our modeling of Stokes V profiles in Sect. 4.2 is that the large-scale surface magnetic field of EK Eri is fully poloidal. This could well be the remnant of the dipolar field of an Ap star: in this case, the strong field observed at phase 0 could be related to a single magnetic pole. In the dynamo context, fully poloidal large-scale magnetic fields were observed in solar twins (Petit et al. 2008) and in completely convective dwarfs (Morin et al. 2008a, b): in this context, the large variations observed in our B$_l$ measurements would be due to magnetic spots. Only complete modeling and knowledge of global geometry and topology of the magnetic field will enable firm conclusions to be drawn.

\section{Conclusion}
In Sect. 3., using bibliographic data and recent evolution models, 
we supported the view that EK Eri is a 
subgiant star that has just crossed the Hertzsprung gap, 
and has a rotation period of 306.9 d. Our direct 
detection and measurement of the large scale magnetic field of EK Eri indicates that this field is as strong as that observed in RS CVn or FK Com type stars and far 
stronger than observed for other  slowly rotating 
main-sequence stars or in any single giant star observed until now with NARVAL.  Our 
magnetic observations demonstrate the presence of a strong large-scale magnetic field at the surface of EK Eri, which is  unexpected for a slow rotator in the context of a solar-type dynamo.
This situation is consistent with the suggestion of  Stepie\'n (1993) and Strassmeier et al. (1999) that EK Eri could be the descendant of a strongly magnetic 
Ap star. Clearly, a magnetic survey of EK Eri for a few rotation periods is necessary to map the surface of the star and study 
possible variations in its magnetic structure and determine its true origin.

Magnetic Ap stars represent about  $5\%$ of A-type  main-sequence stars, which are among the parents of the G-K evolved stars. The observation of other 
descendants of Ap stars during the red-giant phase can now be addressed: but where are they? Koen et al. (2001) discussed another evolved Ap star candidate, 
HD 21190, possibly an F2III SrEuSi. In this evolutionary state, convection has not yet reached sufficient depths inside the star for the surface abundance anomalies, which will 
disappear in the G phase, to be removed, as in EK Eri. Some X-ray active giants in the Gondoin (1999) sample had small $v \sin i$, and some 
stars with variable Ca~{\sc ii} H \& K emission had periods longer than 100 days (Choi et al. 1995). The classical solar-type ($\alpha \omega$) dynamo may be 
too inefficient to maintain moderate magnetic activity in these stars, and the presence of fossil magnetic fields from weaker magnetic Ap stars may be invoked.
Furthermore, Charbonnel \& Zahn (2007b) suggested that the descendants of Ap stars hosting a strong 
fossil magnetic field should escape 
the thermohaline mixing that occurs at the bump of the red-giant branch 
in $\sim 95 \%$ of low-mass stars (Charbonnel \& Zahn 2007a). 
They could be recognized as corresponding to abundance anomalies in Li and $^{12}C/^{13}C$ 
with respect to the non-magnetic stars. This criteria cannot be used for EK Eri, since 
it has not yet reached the luminosity bump where thermohaline mixing becomes efficient 
in non-magnetic giants.
A sample of these stars will be observed with NARVAL and ESPaDOnS, and our understanding 
of the fate of the 
fossil magnetic fields during the red giant branch will be significantly improved in the near future.

\begin{acknowledgements}
       We thank  the TBL team for providing service observing and help during the observations.
\end{acknowledgements}

\end{document}